\documentclass[lettersize,journal]{IEEEtran}
\usepackage{amsmath,amsfonts}
\usepackage{algorithmic}
\usepackage{algorithm}
\usepackage{array}
\usepackage[caption=false,font=normalsize,labelfont=sf,textfont=sf]{subfig}
\usepackage{textcomp}
\usepackage{stfloats}
\usepackage{url}
\usepackage{verbatim}
\usepackage{graphicx}
\usepackage{cite}
\usepackage{caption}
\usepackage{booktabs}
\usepackage{hyperref}

\usepackage{color,soul}

\begin{document}

\title{Infrastructure Engineering: A Still Missing, Undervalued Role in the Research Ecosystem}

\author{Vanessa Sochat \\~\IEEEmembership{Lawrence Livermore National Laboratory}
\thanks{Livermore, CA}
\thanks{Manuscript received April 19, 2021; revised August 16, 2021.}}

\markboth{Journal of \LaTeX\ Class Files,~Vol.~14, No.~8, August~2021}%
{Shell \MakeLowercase{\textit{et al.}}: A Sample Article Using IEEEtran.cls for IEEE Journals}


\maketitle

\begin{abstract}
Research has become increasingly reliant on software, serving as the driving force behind bioinformatics, high performance computing, physics, machine learning and artificial intelligence, to name a few. While substantial progress has been made in advocating for the research software engineer, a kind of software engineer that typically works directly on software and associated assets that go into research, little attention has been placed on the workforce behind research infrastructure and innovation, namely compilers and compatibility tool development, orchestration and scheduling infrastructure, developer environments, container technologies, and workflow managers. As economic incentives are moving toward different models of cloud computing and innovating is required to develop new paradigms that represent the best of both worlds, an effort called ``converged computing," the need for such a role is not just ideal, but essential for the continued success of science. While scattered staff in non-traditional roles have found time to work on some facets of this space, the lack of a larger workforce and incentive to support it has led to the scientific community falling behind. In this article we will highlight the importance of this missing layer, providing examples of how a missing role of infrastructure engineer has led to inefficiencies in the interoperability, portability, and reproducibility of science. We suggest that an inability to allocate, provide resources for, and sustain individuals to work explicitly on these technologies could lead to possible futures that are sub-optimal for the continued success of our scientific communities.
\end{abstract}


\section{Introduction}

When we think of research we often think of software. Research software is now a driving force behind work in bioinformatics \cite{Noor2022-ut,Wratten2021-iz}, machine learning and artificial intelligence \cite{Lo2021-kz}, chemistry, systems biology and medicine \cite{Keith2021-jz,Mesko2020-wd}, and a gamut of other fields \cite{Jay2020-iw}. While there still exists a divide between wet and dry labs, conducting research with meaningful analysis is not possible without software, whether it is needed solely for an analysis, processing of data, or integrated deeply into the science itself. We call this software that is used directly for scientific research to record data, run analyses, or other domain-specific tasks scientific software, and entire communities have grown from it \cite{Carver2022-rw,Sochat2022-ac}.

Historically, the responsibility for developing scientific software fell on the scientists themselves. In the last decade, the role of Research Software Engineer (RSE) has emerged to fill that gap, where traditional RSEs are those that work on research software from within a lab, alongside a research computing support group, or as researchers themselves \cite{Hettrick_undated-sl}. While this movement has led to improvements for scientific software that directly supports research, a primarily academic incentive structure, one based on short term publication, presents significant issue with both the longevity and innovation of research infrastructure. The main issues arises because foundational libraries and underlying infrastructure are considered supplementary to and in support of science, but not necessary science themselves \cite{Howison2011-hq}, and thus are not a focus for innovation until there is a problem that forces attention to them. Ironically, this underlying infrastructure that ranges from system software to container technologies and scheduling infrastructure is so immensely important that the scientific ecosystem would not function without them \cite{Carver2022-rw}. 

An understanding of the need for infrastructure engineers can come from the history of software development in science. Container technologies \cite{Kurtzer2017-xj,Priedhorsky2021-xx,Gerhardt2017-tf} and workflow tools provide key examples of reactive solutions, or the larger community pushing for innovation in a software space to solve an immediate problem instead of taking preventative approaches. This need for quick fixes, often pushed onto small and understaffed research support teams, makes it increasingly difficult for the scientific community to keep up from the perspective of infrastructure innovation. A persistent and well-staffed layer of infrastructure engineers doing the work and research to keep up with the latest trends and innovate in the space is badly needed to avoid this reactionary software development. 

The need for infrastructure engineering is further compounded by the emerging economic powerhouse of the cloud, 
which is expected to reach over 1.1 trillion dollars of revenue by 2027 \cite{hyperion-bloob}. Given the inordinate profits and ability to hire and develop hardware and software in-house, cloud companies are driving the space of innovation for compute, storage, and consequently, paradigms and underlying infrastructure that support it. In that the scientific community also has come to rely on some of these resources and minimally could be impacted by changes in the hardware and software space, this is cause for concern. For example, a choice to develop and primarily provide lower precision chips \cite{Agrawal2021-ze,Lew2023-ry} to support the artificial intelligence (AI) market would impact researchers that rely on high precision chips \cite{Adam_Paxton2022-ok} for their models. Scientific simulations that rely on communication frameworks that warrant low latency networks \cite{Underwood2018-jp} will not function well if these environments have developed and use a different paradigm. A lack of understanding to the costs and ability to compare with on premises resources will further hinder a researcher's ability to make sound economic decisions \cite{pauley2010cloud}. If cloud encompasses a large share of the compute market and the scientific community will eventually be absorbed into that space, science will suffer if preparatory work and experimentation is not done to understand performance of current HPC applications and innovate converged or new paradigms in that space. This possible future where applications and simulations are not able to run on the compute resources available and scientific progress slows or stops is not desirable.

In this paper, we provide background about research and development of infrastructure, and how different roles have been involved. We start with a discussion on incentives (Section \ref{section:incentives}), and suggest that the historical and current academic model does not provide sufficient incentive for the work needed for the continued success of science. We then trace key projects such as container technologies and the Message Passing Interface (MPI) (Section \ref{section:reactive}) as examples challenges in the infrastructure space. We delve into the exploding domain of cloud computing, and speculate on how the economic landscape might impact high performance computing and science (Section \ref{section:transparency}). By way of introspecting on the space of incentives that drive innovation of scientific infrastructure and discussing current challenges, a set of possible futures in the research community can be considered. Thus, this paper focuses on current challenges in the space of innovating software and infrastructure for science, and provides insights to past events. Notably, the best possible future for our scientific work requires this additional focus on core technologies and infrastructure, a goal that is complementary to but often unrelated to the pursuit of developing research software.
\section{Infrastructure Development}

\subsection{Incentives for innovation from academia}
\label{section:incentives}

\begin{figure}[H]
  \centering
  \includegraphics[scale=0.25]{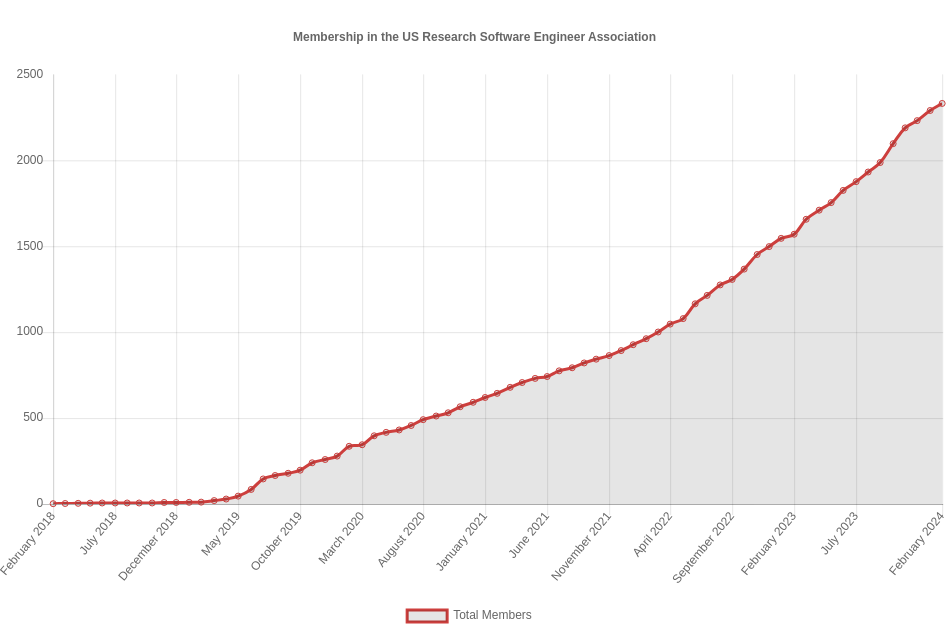}
  \caption{Membership in the United States Research Software Engineering Association. Despite its recent foundation, the official membership count has grown steadily.}
  \label{fig:usrse}
\end{figure}

Scientific software came before infrastructure to enable it. Historically, the responsibility for developing scientific software fell on the scientists themselves. An early example of early software development being connected to scientific research comes from Lorenz \cite{Lorenz1963-ez}. His work not only investigated a scientific question related to weather prediction, but the practice and method of answering the question itself. While software has been prominent in research since that time, with 90-95\% of researchers directly relying on it \cite{Carver2022-rw}, it was not until 2010 that an empirical study \cite{Prause2010-nc} identified the trend and coined the term ``Research Software Engineering." It followed logically that the individuals working on research software were ``Research Software Engineers" and a movement to more formally define and support the role started in the EU in 2012 \cite{Hettrick_undated-sl}.  The role of the research software engineer (RSE) has thus emerged in the last decade, where traditional RSEs are those that work on research software from within a lab, alongside a research computing support group, or as researchers themselves. Growth can be seen on the level of countries or from within. The movement started in 2012 in the United Kingdom \cite{noauthor_2019-zx} and has since grown to include the United States \cite{Sochat-usrse}, Australia \cite{RSE_Australia_New_Zealand_RSE-AUNZ_undated-mb}, Asia \cite{RSE_Asia_Association_undated-od} and other geographic areas. Figure \ref{fig:usrse} shows the quick expansion of a community within a country, the United States, growing to almost 2.5k members in under 5 years. Growing out of the academic community, the community itself has inherited an academic incentive structure, with a large emphasis of value placed on publication, conferences, and integration into the scientific grant system \cite{noauthor_2020-my,noauthor_undated-to}.  This inheritance of a publication-based model as a primary means for visibility means that research software engineering groups also strive for publication to demonstrate the value of the software. As a consequence, many roles for research software engineering still prefer candidates with advanced degrees in scientific disciplines, and there is a strong expectation for continued participation in the publication process. This is reflected in the emergence of journals that are explicitly for research software \cite{smith2018journal,jors,derrick2021assessing}. While it is unknown if the RSE community will take on the work to explore alternative models of valuation, the inheritance of the traditional academic publication model that requires quick software development to solve a specific need toward publication is in opposition to one based on longer term, persistent research to continually improve upon an ecosystem of infrastructure that lies outside the laurels of scientific accolade.

This underlying publication-based incentive structure presents problems for the longevity of software projects. A small study \cite{VanessaSaurus2023-wm} using data from the Research Software Encyclopedia database \cite{Sochat2022-ac} demonstrated an alarming trend that much of published research software ceases development soon after publication. For a set of approximately 5000 software projects, after 14 months only 20\% continued to be worked on, as evidenced by activity in GitHub. Extended out to 40 months, the set was reduced to only 63 projects. The author called this set ``high value" projects and postulated that if a piece of software is still active long after publication, it is likely a reflection of being important for the research ecosystem. Interestingly, the high value projects did not tend to fall within traditional research software that has a domain science focus, but rather included package managers, workflow tools, languages, plotting libraries, machine learning and statistical libraries, and core numerical libraries. Many of these projects also were paired with sustainable funding, with over 50\% having primary backing from a foundation (e.g., NumFocus or Sloan), a grant, or national laboratory (Table \ref{tab:funding}). Interestingly, the foundations themselves (e.g., NumFocus) tend to be corporate sponsored, suggesting that much software sustainability ultimately comes from the private sector. Whether the funding came first or the value of the project is a chicken and egg problem. It can be postulated that projects that emerge in the research ecosystem build up a community of contributors, and the contributors seek out sustainable funding sources after reaching a threshold of contribution and value. It might also be the case that industry partners come to rely on a piece of software, and encourage action to directly support it. However, short term longevity does not indicate continued persistence. Since that work, several (approximately 6) of the projects have not had significant activity, a reflection of the continued struggle for sustainable research software. 

Case studies \cite{Howison2011-hq} of large scientific, collaborative efforts in 
the fields of microbiology, high energy physics and structural biology tell a story of similar incentives -- that instrument and software development was done in the pursuit of science, but not as a career. This ``service work" in fact, might lead scientists too far down a development path and serve as a hindrance to a successful academic career, which requires citations to grow reputation. The software is viewed as supplementary or a service, and the work on it often temporary, expiring with the publication and typically end of grant funding. Maintaining any software beyond this time, aside from the monetary and staff burden, would directly conflict with pursuing new work toward continued success of an academic career \cite{Howison2011-hq}.

The observation that valued software goes beyond traditional domain or academic software in scope is supported by similar and ongoing work \cite{VanessaSaurus2022-ty,Sochat2022Citelang} to generate a credit score not only for top level citations of software, but also underlying dependencies. The work found that, across languages with prominent representation in the database (R, Python), the most highly scored software in the database was  libraries for data formats and parsing, numerical analysis and plotting, testing and documentation. These are foundational libraries that make up research software, but are not considered research software in and of themselves. From an academic standpoint, these libraries are not the ones that are published by the scientific community because they are considered material to science and do not directly conduct research. As the lowest level of dependency, they are often not cited, and might be taken for granted. Ironically, the scientific ecosystem would not function without them. This work demonstrates there can be a mismatch between what is needed and what is superficially valued, or brought to the forefront of awareness via publication and publicity.

\begin{table}[H]
  \centering
\caption{High Value Research Software Funding}
  \begin{tabular}{lll}
    \toprule
    \cmidrule(r){1-2}
    Source     & Count  & Percentage    \\
    \midrule
    Foundation & 42 & 51.0\%   \\
    Open Source & 24 & 29.0\%   \\
    Industry & 11 & 13.3\%   \\
    Government or Research Institute & 6 & 7.2\%   \\
    Sponsors & 4 & 5.1\%   \\
    \bottomrule
  \end{tabular}
  \label{tab:funding}
\end{table}


While the success of the Research Software Engineering movement is an accomplishment for the role of RSEs, the academic incentive structure remains a significant problem for the interoperability, portability, and reproducibility of scientific workflows. Unless the software is directly included as an entry in the database, the work mentioned previously cannot account for infrastructure and core software libraries that are ingrained in the scientific process.  This might include (but is not limited to) workload managers, version control systems, integrated development environments (IDEs), drivers for hardware, operating systems, compilers, and testing libraries. The Research Software Encyclopedia calls this software ``incidentally used for research" \cite{Sochat2022-ac} and suggests that although it is unlikely to be published or cited, it is core to the day-to-day functioning of science. As an example, container technologies extend beyond research software in that their development often requires manipulating namespaces, and as an isolated unit, their innovation would be harder to publish to tie to a successful academic career. Yet they are undeniably an essential component of the research ecosystem, with almost 80\% of the HPC community using them at least once a week \cite{sochat_2024_11206333}. Arguably any of these hidden infrastructure projects might be worked on and optimized from a research perspective. From the practices stated here, it is suggested that traditional academic incentives are not sufficient to support innovation of core research infrastructure. For the purposes of this paper, we will label this layer of software as infrastructure, and the work that encompasses it infrastructure engineering. 

\subsection{Incentives for innovation from computing providers}
\label{section:incentives-computing}

High performance computing (HPC) centers exist to provide a service to run scientific software \cite{Ajani2024-zi}, responding to the needs of society or taxpayers to maximize science, with large grants being spent on bringing up new, larger clusters that strive to efficiently use resources for scientific simulations and scaled discovery \cite{Goponenko2020-qi}. The need for reliable, consistent availability typically means that conservative approaches are taken for the design of the systems and software, and time and staff is often not available for innovative, risky work. While cloud computing was originally considered complementary to on premises computing, the emerging economic powerhouse of cloud resources that are now driving development of underlying hardware and software is now (culturally speaking) considered more of a competitor \cite{Reed2022-sj}. 

For a point of comparison, the exploding space of cloud vendors \cite{Achar2020-xa}, arguably also taking a similar path of providing compute and resources around it, are responding to the needs of their consumer market to maximize total addressable spend (TAS) \cite{noauthor_2023-dd}. Given the huge monetary backing for cloud \cite{hyperion-bloob}, there are often no barriers to staffing or paths that can be explored, and so addressing these needs can delve into bleeding edge ideas and technologies. Companies that pop up to offer "science as a service`` also respond to the needs of a market, where the market might actually trickle in from scientific groups that can afford the services \cite{Sunyaev2020-wi}. Thus, while cloud companies have a strong, market-driven incentive to create novel products to maximize revenue and minimize costs, traditional compute centers that exist alongside academic centers do not. In fact, due to the large user base in a multi-tenant environment, change or moving quickly is seen as more risky, and a conservative strategy more ideal. This lack of incentive to optimize to save costs can be seen in the poor utilization of power for large centers \cite{Patki2021-hg} in comparison to vendor efforts toward environmental friendly practices \cite{noauthor_undated-mv}. To make an analogy with temperature, cloud vendors are monetarily ``exothermic'' while HPC is ``endothermic'' \cite{Reed2022-sj}. Between these two models, the latter is far more challenging to maintain over the long term.

With this reality that cloud is becoming not just a leader, but an innovator in the space \cite{Thompson2021-ww,Reed2023-id}, some have started to imagine a future oriented toward converged computing \cite{Misale2022-nc}, or bringing together the best of both worlds between cloud and high performance computing. This work not only tests the performance and feasibility of the current needs of the HPC community in cloud environments, but also imagines new paradigms and designs that might integrate or allow for co-existence of the best features of currently disparate communities \cite{Sochat2023-us,Misale2021-tv}. To fully support this work, infrastructure engineers are needed to not only work but \textit{focus} on the task of imagining futures for computational paradigms that don't exist yet. These infrastructure engineers need to do foundational work to establish new paradigms for research, and do so in a way that meets the research community where they are at.

This set of infrastructure engineers that might not only develop core, underlying technologies for research but also exist as a glue to bridge communities, largely are missing or accidental. Developer teams oriented toward innovation and explicit work on underlying models for practicing research don't typically exist in traditional HPC centers. The reason is likely simple -- there often is not bandwidth or funding, and the work is  not addressed until there is a reported problem. Most scientists rely on large, community-based software projects, and few frequently rely on high performance computing \cite{Hannay2009-ki} and thus are further lacking incentives to innovate in the space. At best, system administrators and support staff might collaborate with research groups to slowly and carefully test new ideas in rare amounts of free time. The lack of structure, time, and funding for this work is not acceptable given the quantity and speed of change that is warranted.  In the same way that the burden of working on scientific software has historically fallen on scientists \cite{Milewicz2019-pn} and this has led to issues with reproducibility and scientific integrity, the endothermic nature of HPC sustainability \cite{Reed2022-sj} and pressure to provide consistent investment in new systems to afford publication \cite{Apon2010-gg} paints a picture of the HPC community investing the majority of time and money for faster, newer systems with a streamlined group of administrative support staff over anything else. However, with the changing landscape and growing need for new paradigms of compute, full time hires are needed to work explicitly on these problems.

To tie together the story of computing providers between cloud (industry) and high performance computing (scientific research) the current state of the world presents with two disparate communities. From conferences (\href{https://events.linuxfoundation.org/kubecon-cloudnativecon-north-america/}{Kubecon} vs. \href{https://supercomputing.org/}{Supercomputing}) to means of communication (publication vs. marketing) there is often a missing bridge of communication. A layer of persistent infrastructure engineers, by way of seeking innovation in their respective space, would require an understanding across the technology space and bridge the community gap. An infrastructure engineer employed in a scientific computing context would be able to do the research to understand the performance of HPC applications in a cloud environment, and work to present convincing arguments to the cloud community to take interest in paradigms familiar to them, or recognize an opportunity for new paradigms for their community.  Opportunities to work together would solve problems on both sides. For example, a technology or practice from HPC might offer more performant means to accomplish work, or algorithms for scheduling or scaling that can more effectively assigning resources. This makes it clear that the role of infrastructure engineer extends beyond technical and into cultural and community. From this standpoint, it can be argued that computing providers on both sides might have the incentive to innovate, but on the side of high performance computing, the staffing and monetary incentives are not present.

\subsection{Opportunities for collaboration}
\label{section:incentives-computing}

The degree to which two traditionally disparate communities can work together depends on having an overlapping set of incentives, which often can happen accidentally or opportunistically, and right now an opportunity for collaboration is presenting itself. A recent development \cite{CNCF_Cloud_Native_Computing_Foundation2023-wv} that has been made possible thanks to the growing desire for batch, is the need to run machine learning-oriented workloads in the main workload orchestration tool, Kubernetes \cite{Honeypot2022-yv}. While traditionally created for stateless provisioning of services, Kubernetes, for the first time, was desired to be used as a platform for jobs that would have states -- starting, running, and completing. The high performance computing community, already having interest in using cloud for their work \cite{Hemsoth_undated-ld}, also took interest \cite{Beltre2019-mr} in Kubernetes, possibly with desire to run HPC applications there, or to explore it as a new cloud scheduler \cite{Reuther2018-uj}. Beyond having a desire for basic compute, often through bare metal or virtual machines, this was the first time there was a strong, compelling overlap in the needs of the technology space for cloud and high performance computing. It presented an opportunity for collaboration through shared incentives, and the success of this collaboration largely depends on the ability of both sides to show up.

Despite the disparity between monetary resources, the challenges to provide modern interfaces for monitoring compute, automation of building and testing setups, and reproducible workflows are shared between the two communities. A remaining challenge to this collaboration is is pace of work. The two communities also move at different paces, with scientific development moving carefully and slowly, and industry moving as quickly as possible, often backed by much greater monetary support.

\subsection{Reactionary Development Practices}
\label{section:reactive}

Infrastructure engineering in the academic space typically arises from a place of need or external forces. As an example, container technologies \cite{Kurtzer2017-xj} came only after a multi-year, huge push from scientists to HPC centers to provide an alternative solution to Docker \cite{Merkel2014-da}, and this is reflected by the majority of leading scientific container technologies being born out of national laboratories that had the scientific staff and funding to work on them \cite{Kurtzer2017-xj,Priedhorsky2021-xx,Gerhardt2017-tf}. From a cultural standpoint, the supporting staff around the provisioning of infrastructure are in a second tier role. They are in service to the academics that are the primary drivers of science and backbone of an academic institution. It could be inferred that the relatively lower value placed on software that does not directly contribute to research, perhaps worked on by this level of staff, emerged from this academic hierarchy. Software and infrastructure engineers have not been valued in the traditional academic fabric because they are viewed as materials for research and not research. In the same way that students expect pencil and paper to be provided to go to school, principal investigators and academic staff expect compute resources and paradigms to be available to them.

Workflow tools have also largely developed starting from a dire need, and reactionary to a changing compute landscape. The change started in the bio-sciences \cite{Molder2021-rq,Di_Tommaso2017-lw,8621141}, where scientists were presented with many different environments and needed to run pipelines to do their work. The point is subtle, but it must be noted that these tools were developed not for a direct contribution to answering a scientific question, but because the practice of science would not be possible without them. Especially for these workflow tools that came about in the 2010s, the challenge that presented was needing to run on local HPC workload managers plus an emerging cloud environment that provided virtual machines. The projects have been sustained due to the importance in their respective communities, with many now being backed by companies and foundational support. This innovation was largely reactive, and in response to a problem. Under this model, any specific technology space in science will likely lag behind innovation that happens outside of it. To make an analogy with the medical field, what would be needed  is not to react to current problems or treat symptoms after the fact, but to practice preventative medicine. In infrastructure engineering this means actively staffing and working on traditionally undervalued pieces of research infrastructure that are quietly supporting science.
\section{Challenge Landscape}

\subsection{Conflicting Paradigms}
\label{section:mpi}

The core of the problem that presents a barrier to easy collaboration or integration is that high performance computing and artificial intelligence rely on different paradigms of compute to do their work. As an example, the Message Passing Interface (MPI) is the long standing bread and butter programming model for communication in HPC. MPI was created to provide a low latency means to solve partial differential equations at scale, and thus requires low latency network fabrics. It is used in over 90\% of Exascale proxy applications \cite{Sultana2021-vy} for over three decades \cite{Walker1992-kr}, and applications at leadership facilities can spend upwards of 40\% of machine time using it \cite{Liu2020-uk}. In practice, this led to a desire for HPC practitioners to innovate the underlying hardware for MPI, or address issues or problems that came up when using it over exploring new communication paradigms. While artificial intelligence (AI) could approximate solutions to the same problems using different approaches, machine learning and statistics, what prevents this adoption is the degree of unknowns. It's not clear what problems AI is able to solve, and what approaches are suited for specific problems. The work is poorly understood, and thus not easy to adopt or would take years of research to demonstrate its validity \cite{Li2020-ka}. 




In that work for validation and verification requires inspection of every layer of the stack and can take years, this makes it less likely to pursue. There is also an interesting space to explore that might afford combination of both paradigms. For example, artificial intelligence might provide a faster preliminary model that then might be validated by more traditional numerical approaches, thus accelerating science by speeding up an initial step of hypothesis generation. There is also a missing piece of work to categorize the types of HPC applications that are run on premises, and better understand which are amenable to these converged or cloud-native approaches. As an example, leadership facilities \cite{Liu2020-uk} are optimized for scaled simulation and modeling, and although they can support these workloads, in practice less than 1\% of jobs might use half of the resources offered by a center, which translates to approximately 10\% of core hours \cite{Liu2020-uk}.

While large, multi-physics simulations that use a large percentage of a center's entire resources are unlikely to be suitable for a cloud deployment in the near future, smaller ensemble workloads might be, and herein lies the opportunity to evaluate the paradigms used by these workloads. Instead of an ``all or nothing" perspective to cloud and HPC, a more productive mindset is to consider HPC as a scientific instrument \cite{Reed2022-sj}, and be open to the other that many of the workloads we are currently running on premises could be possible in cloud with more thinking about paradigms. Discovery of new or alternate paradigms would need to come with an open mindset to refactor and port legacy applications to use them.

In summary, the current challenge of the scientific community is a lack of validated approaches that use new cloud programming paradigms, and the mathematical structure of current applications and simulations that make them rigid to software changes. There is also poor understanding of the cost effectiveness and comparison between the two paradigms, and a general observation \cite{Carlyle2010-ga} that suggests that cloud is more expensive. These observations make it less likely for the HPC community to seek innovation in the space without substantial problems arising that force it. 
 
\subsection{Transparency of resources}
\label{section:transparency}

So far, this paper has not delved into cultural challenges beyond incentives, however a major one exists with respect to transparency. Notably in the space of converged computing, there are two misconceptions about resources and cost. Notably:

\begin{itemize}
\item Cloud gives the impression of infinite resources.
\item HPC gives the impression of free resources.
\end{itemize}

Both of these points are not true -- resources in cloud are not infinite, and using HPC is not free. Work is needed that likely would come from this space of developers to provide transparency to these misconceptions, namely doing work to understand the true availability of resources (if cloud vendors are not willing to offer the information) and better understanding the true power and utilization of an HPC center to make informed decisions of not just how to operate locally, but the true monetary trade-off of running on premises vs. on the cloud.

While cloud vendors provide consumers with an ability to see exact prices down to the resource unit, true cost transparency \cite{Hayes_undated-tv} would also require an understanding of how those prices are derived. The issue with a cloud market that lacks price transparency is that consumers have no means to evaluate the quality of the resources they are given \cite{Akerlof1970-tl}. Given two goods, such a storage resource, that are priced equivalently but one is actually of lower quality, the more valuable one will eventually disappear from circulation. An example comes from solid state drives (SSDs) where it was noticed that the advances in throughput were not reflected in SSDs offered by the cloud \cite{Leis_undated-ja}. Whether the reason is due to providing older equipment or purposefully setting caps cannot be known. In this scenario, two clouds might both offer SSDs at comparable, outwardly competitive prices, but the underlying devices might vary hugely in the throughput. In the long term, the cloud offering the faster devices would have no incentive to continue doing so, and the consumer would ultimately be hurt having only the option for slower storage at a price that does not reflect the lower value of the good. What is needed for full transparency is an ability to trace the full rationale behind a price point, likely deriving the cost of a unit of storage or compute from a combination of commodities and engineering costs, and value by providing it as an on-demand service.

A second common complaint about using cloud is with respect to cost planning. While there are dark patterns that arguably could be pointed out and resolved (for example, the cost of waiting for a node pool to be deployed), it cannot be overlooked that clouds provide good information about the prices they charge, down to dollar charges per unit of usage. What is needed to address these undesirable events are tools or protocol taken to use that information in a meaningful way. Infrastructure engineers would be prime to develop such tools. There is also an element of collaboration needed to address the dark patterns. What is needed is communication over accusation -- having a discussion about the issues with vendors, bringing the issues to light, and working on solutions. Generally speaking, cloud vendors are companies that care about customer feedback, and given the desire to create a positive brand and attract customers, will be appreciative and likely responsive to feedback from enough of them. It is this missing layer of infrastructure engineers studying power, performance, and utilization, that can take on the bulk of this work.

HPC also struggles with transparency in that it is rare to find openly accessible models of cost for centers.
This is a challenging ask. Centers have different definitions of utilization that range from the number of nodes being used to percentage of cores or other resources, to (less likely) actual estimates of power by way of floating-point operations per second (FLOPS). From the user perspective, most of this information is not shared publicly and there is an impression of free resources, and at most, some scientific groups are asked to request reservations for scoped works that come with a monetary cost or unit. These same infrastructure engineers would be in a prime position to explore common units for cost, possibly based on better understanding of power and utilization, and to pursue units that might be mapped across institutions and then compared with cloud. Notably, it would be interesting to see plots of mapping capability against cost effectiveness, and taking into account not just utilization of resources but also people. The challenge for the community will be to maintain capability while increasing cost effectiveness. In a larger sense, not having staff that can quantify resource usage and then explore these ideas will ensure that the evaluation of costs between cloud and HPC remains muddled and confusing at best.



\subsection{Community Participation}
\label{section:containers}

Participating in collaborative discussion around the development of new paradigms is important to have influence on their development. This is especially important for technologies that might touch communities with different needs that might drive different development paths. When this does not happen, it is often the result of lacking time, interest, or incentive. An early example of a bridging technology comes from the container technologies space. The first container technology, Docker \cite{Merkel2014-da} was demoed at an industry conference in 2013. It was a taste of the future to come, offering seamless environments for applications and development that leveraged kernel namespaces. This technology, broadly referred to as Linux containers, became a new unit of development that streamlined productivity, and naturally it came with a desire to provide registries \cite{Cook2017-lq}, runtime implementations \cite{Wang2022-zd}, and standards to support that. In 2015 the Open Containers Initiative (OCI) was founded, reflecting a need for standardization. It could not be the case that every company using container technologies would develop a slightly different means to run containers, pull or push them from a remote collection called a registry, or describe their metadata or interactions. This is the foundation of the basic need for standardization, so that software can work across environments with minimal friction \cite{noauthor_undated-fa}. 

It would have been ideal for members for the high performance computing community to be present at these early meetings to advocate for the scientific or HPC use case. While a handful of scientific community members  showed up in scattered increments, it can easily be seen by looking at the technical oversight board throughout the years that the majority of voices came from industry \cite{noauthor_undated-wa}. As a result, a set of successful standards were developed that were oriented toward industry or cloud use cases, where notably it is less common to see multi-tenancy, and more common to have privileged access to a system. 

Interest from the HPC community to have support for the same technologies \cite{Kurtzer2017-xj} came with a desire to to address the reproducibility crisis \cite{baker2016reproducibility}. However, most of this advocacy was retroactive -- figuring out ways to support such a technology in a space it was not designed for, one with multi-tenancy and constrained user permissions. A lot of time was also spent trying to convince maintainers of clusters that providing such a technology was a sound thing to do from a security standpoint. Arguably, having a voice at the table early on might have made this initial adoption step easier. It could also have been that these early efforts for advocacy of container technologies failed with too many technical challenges to overcome. Thankfully, pockets of staff that predominantly worked in research computing or other national laboratory groups stepped up to the challenge, including the Singularity project out of Lawrence Berkeley National Laboratory \cite{Kurtzer2017-xj}, CharlieCloud out of Los Alamos National Laboratories \cite{Priedhorsky2021-xx}, and Shifter out of NERSC \cite{Gerhardt2017-tf}.

The story above demonstrates that containerization for the scientific community developed in a delayed fashion. It required responding to an existing model that was developed for a different use case, and only when it directly solved a current problem and was pushed for. These early decisions trickle down into later ones, adding more toil for scientific adoption. For example, Kubernetes \cite{Honeypot2022-yv} was developed with the (at the time) rootful Docker container as the main unit of operation, and the design of requiring elevated privileges for most components still persists today. Projects that try and address this issue are small and rare, but still offer an entry point for the scientific community to engage and advocate for their needs. 

In 2022 there was a KEP (Kubernetes Enhancement Proposal) ``Usernetes'' that offered up a design for running Kubernetes in user space. With only a handful of developers, the project did not have a push for development and was in danger of becoming dormant. It was not until an interested party on the HPC side stepped in in 2023 to re-ignite discussion on the need and development of the project did it continue to progress \cite{fosdem-bare-metal-bros}. At the end of 2023, a generation II of Usernetes was released, and one that used more modern approaches with rootless containers and streamlined deployment. With this small push, the project now has the chance for being brought back to life as a means to bring the automation, modularity, and service-oriented approach offered by Kubernetes to user-space environments. This work is novel, exciting, and just beginning to be showcased today \cite{fosdem-bare-metal-bros}.

The example above demonstrates the impact that even a single or few number of staff that are oriented to innovate on behalf of the scientific community can have. Arguably, a larger, and persistent layer across academic centers and national laboratories might have greater impact. The story around convergence of container technologies continues with a new OCI working group focusing on compatibility of container images. By way of some of the same community members, the voice of the scientific community is well-represented, and proposals have emerged that take into account HPC use cases \cite{noauthor_undated-lq}. 

\subsection{Developer environments}

The Usernetes prototype (Section \ref{section:containers}) provides a strong example for a missing and often overlooked component: developer environments. In high performance computing especially, the idea of ``developer experience" is often forgotten, misrepresented or misunderstood. While it is often inferred that ``developer environment'' refers to one thing, there are actually two kinds, and the one that is needed for core infrastructure work and innovation is often forgotten. The most common definition of a developer environment references one on a multi-tenant system, and is an environment for a code team to work on scientific software. It is intended for the research software engineer use case. There are several tools oriented for this use case \cite{Hoste2012-dj,Gamblin2015-uv}, often coming from package managers being able to create isolated views with software, or module systems. These views carry the same limitations as the system they are installed to, primarily not allowing for escalated privileges, and installing assets to a scoped space where the user has permission to write and execute. Infrastructure engineers that are working on container technologies, workflow tools, operators and storage that require permissions not afforded in a multi-tenant HPC environment are often forgotten. As a result, they often primarily develop on local machines or (if lucky enough to have access)  cloud resources. It isn't unheard of for these developers to spend their own money on resources when they cannot be provided by the institution, because their work is impossible without them. Their work is badly needed, and needs to be supported. It often has nothing to do directly with science, but can single-handedly enable it. 

Assuming that a core team of software engineers exists at a center or institution, the ability of these team to, for example, have an environment that can be created and destroyed, with full (root) permissions, must be a priority. The Usernetes example is relevant here, because for the work described in \cite{fosdem-bare-metal-bros} a one-off, custom environment was needed to allow for provisioning custom virtual machines, and allowing them to have features like cgroups2, custom enabled kernel modules, and be external to the core networks and services provided by the core center. This last step is to ensure that the development work does not add any additional danger or risk to the functioning of the primary resources. It must be pointed out that procuring this kind of setup, often through layers of SSH and authentication schemes only in the terminal, can be challenging. If setting up this development environment is not possible,  which is likely the state at most centers, the work would not have been possible, and the incremental step toward a new prototype for research would be nonexistent. It is easy to imagine all of the missing work and problems that might have been solved in the last decade if infrastructure teams were hired solely to work on these core resources, and provided with the developer environments they needed to be productive. These teams minimally need the following two components:

\begin{itemize}
\item \textbf{Portable, and throwaway environments} that can be spun up and used, turned on and off, for the sole purpose of innovating the technology (storage, network, containers, workflows, etc.) that will trickle back into center adoption.
\item \textbf{Modern interfaces} that feel like local environments. E.g., SSH sessions that die, are behind three connections, or don't give a nice (fast) interface with access to all the resources needed are arduous to use.
\end{itemize}

This lack of \textit{infrastructure engineering} developer environments is a key problem in the challenge landscape that must be addressed to empower infrastructure engineers to do their work.

\subsection{Models of convergence}
\label{section:models-convergence}

Perhaps the largest challenge that might be faced toward work to innovate infrastructure to support science is identifying strategies for working together. These are models of convergence -- patterns that allow for successful convergence between these two communities, and often they are both technological and cultural. The first pattern is with respect to shared \textit{incentives}, which have already been talked about in section \ref{section:incentives}. It is going to be less likely that two communities can successfully collaborate if they do not care about similar things. When incentives are aligned, staff and resources can be assigned to tasks from higher level management, and goals defined around solving the problems. 

The second pattern is with respect to \textit{modularity}. The degree to which technologies from different interfaces are designed with modularity in mind can directly influence the ability to swap components in and out of the interface from the opposing community. As an example, the modularity of the Kubernetes scheduler, and a plugin-based architecture \cite{Queue_undated-sb} allows for an interested community member to easily design a plugin that modifies functionality for their needs or experimentation.  This is a huge potential area for collaboration, as the HPC community has huge expertise in scheduling and strategies to offer. With this collaboration, a possible future is that cloud resource provisioning is improved by HPC scheduler technologies \cite{Patki2023-cg,Liu2022-pg,Misale2022-nc}.

A third and very simple design strategy for converged computing is \textit{integration}. This requires creativity to implement the entirety of one component inside of another. As an example from HPC, it has been possible to integrate several workload managers into the Kubernetes space by way of containerization and creative use of Kubernetes abstractions to map needed components into Kubernetes \cite{Sochat2024-xd}.  Similar work has been done for Slurm, although the software is not public \cite{sunk}.

The last strategy for converged computing is \textit{co-existence}. This means optimizing two traditionally different technologies to work together, either side-by-side or with interaction in the same space. The example from this paper that has already been highlighted is Usernetes, or running Kubernetes in user space. This technology has been demonstrated in a prototype \cite{fosdem-bare-metal-bros} to co-exist alongside a bare-metal install of a workload manager, allowing for a subset of components for a heterogeneous machine learning workload (e.g., simulations that warrant the speed of message passing interface with an HPC network) to run alongside services that the workflow needs (e.g., a task queue or machine learning server to send results for training). This new model offers a means to run workloads on HPC that are more similar to what is seen in Kubernetes, and thus provide an easier means for collaboration across those spaces. A problem that remains to be solved is the network setup itself inside of Usernetes, which is often a factor of two or more times slower due to needing to send traffic through an additional tap device \cite{Matsumoto2024-zn}. Community members from HPC that have expertise in systems and networking might consider engagement to help solve this problem.

\section{Possible Futures}

The discussion of the challenge space can now be extended into an exploration of possible futures. For several known problem spaces, the sections below offer perspectives for what work might be afforded by infrastructure engineers to improve outcomes for science. 

\subsection{Workload Reproducibility}
\label{section:reproducibility}

Reproducibility is often presented as a binary unit of truth -- that an application or workflow is fully reproducible, or not. Another perspective is that reproducibility is a dimension of choice that a particular community or group can choose to operate on. For any choice along that continuum, which might come down to bitwise replication of an application run, demonstrating that something can run from start to finish again, or coming to the same conclusions from subtly different data, the choice can be subjective or depend on the person or group defining the metric. Infrastructure engineers can take on the task to define levels of reproducibility important for the communities they serve. For example, if reproducibility relies on provenance of data, then better research and prototyping is warranted for data containers that might not just be pulled and extracted, but mapped and mounted across cluster nodes. If a future of compute on the cloud means lower precision chips, and climate scientists will need to use these resources for some of their work that must be reproducible, research and work is needed to understand how to do that. 

Environments to afford running and reproducing scientific workloads also present challenges to be worked on by infrastructure engineers. A future can be imagined where HPC resources are intelligently organized with respect to storage and software, and using standard formats and automation to interact with all of the above. A lack of standards for organization and formats badly hurts not just external reproducibility (e.g., sharing work with collaborators) but internal reproducibility too, which is being able to reproduce running a workflow on the same or similar resources at a later time point. To address this issue, infrastructure engineers might test several new paradigms. The first for distribution of software is the often undesired mono-repository approach, which is utilized by many large tech companies to provide a single source of truth \cite{Jaspan2018-em}. A team of engineers with dual build and support expertise would be required to enable this. 

Another approach to test is a modular one, provisioning environments on demand from reproducible build specifications, giving code teams exactly what they need when they need it for a development session. This setup would need to allow saving an active environment state for later if needed. Importantly, the step of giving people exactly what they need at frequent points in time forces them to think about the environment, request it, and then capture provenance about the request to reproduce it. This approach also allows for flexibility of what software, environment bases, and versions are supported. HPC often uses environment modules \cite{Na2020-ui} to support many different versions of software, and often requires manual requests to update to new versions. This on-demand environment approach could provide an intermediate between a mono-repository and modules approach. An explicit choice can be made by maintainers of the on-demand environments about deprecation of software versions in favor of updates that often come with security fixes and improvements. While no code team wants to hear that they must update their code with changes to support a new version, if software is imagined as a living entity, it follows that its survival is contingent upon these same updates. While it adds extra work for code teams to update, the frequent updates arguably could save time down the line when an old code base needs to be renewed, and several months or years of updates need to be made.

\subsection{Performance and Utilization}
\label{section:performance}

Environments for planning and measuring attributes of software components are also needed. Right now, running jobs and requesting resources is at the manual decision of the running user, and this leads to inefficiencies such as asking for incorrect resources, or running a large batch job, finding there is a mistake, and being forced to schedule and run it again. Much of this inefficiency arguably comes from a lack of complete understanding about application design patterns, resource utilization needed for different designs, and thus optimal scheduling. Without this understanding, the result is erroneous runs that either clog a system and add further delay the scientific process, or runs that utilize resources inefficiently. 

Infrastructure engineers are primed to tackle this problem from two angles. The first is identifying design patterns and associated performance trends. For example, a CPU-bound application that spends only a small time in communication might still work well without an extremely low latency network. The second step is to provide tooling for automated understanding of the discovered design patterns or needs. In practice, for the scientific user this might manifest in doing a test run before scaled deployment to collect this information. They  could then have higher confidence for the successive runs to perform optimally at scale. Tools already exist for profiling the needs for storage and IO \cite{Xu2023-ro}, for CPU and power utilization \cite{Patki2021-hg}, and a myriad of other factors. Another low level tool for performance analysis that warrants further exploration is the Extended Berkeley Packet Filter \cite{vieira2020fast} (eBPF), which has taken off for tooling in the cloud community. Infrastructure engineers might better explore  this technology for the HPC use case.

We can extend this vision of profiling and performance design to one that improves upon scheduling of workloads. Profiling information can directly be integrated into a scheduler to optimally select resources, and in fact, the compatibility metadata mentioned previously \ref{section:containers} can be the vehicle to deliver it, a consistent standard artifact used by both cloud and HPC. With optimized means to test, run, deploy, and operate,the entire research ecosystem could be made more efficient, reproducible, and pleasant to use.

\subsection{Interoperability and Portability}
\label{section:portability}

A discussion of possible futures for interoperability and portability, for both software and data, starts with container technologies. Despite containers being relatively popular in the HPC space, it is still immensely challenging to containerize a new application. Arguably, this should be easy to do, and in the process, capture patterns of needs for systems (e.g., network, jobs, IO/storage, compute, etc.), and understand how those patterns map between spaces. While research software engineers can help with building containers, infrastructure engineers are needed to work on the underlying tools, registries, and standards that are used for building and provisioning containers.  These infrastructure engineers would likely be developing and contributing to core container technologies that are used by researchers. This is a multi-directional process. The needs of the HPC or scientific communities first need to trickle up to proposed solutions to the software directly, and as a second step, when some consensus is reached, discussion is followed up by contributions to code. However, there is a third step. When a solution is implemented, it is not enough to have it exist in silence. The same developers that championed the approach and contributed to the change need to turn back and discuss the usage with the research software engineers and scientists. It becomes not just a technical question of development, but one of advocacy and education.

To supplement the above possible future where building containers is not arduous, we can imagine a more streamlined process to building, provisioning, and running containers. Such a process would lead to fewer issues with running them, and free up system administrator time to work on more advanced features or side projects. This could lead to a cultural shift where it's easier to practice science and develop and innovate tools that use containers as units of operation.

Interoperability transitions next into workflows. Workflows present an interesting story because they emerged in the high-performance computing and scientific communities first. Popular workflow tools \cite{Molder2021-rq,larsonneur2018evaluating,Di_Tommaso2017-lw} have embraced a common design pattern of using an executor backend, where it's possible to define a common unit of work, the work is mapped into a directed acyclic graph (DAG), and then based on the executor, submit to different environments to run. In the early days this selection of executors included modes for local and workload manager execution, and as clouds have improved their application programming interfaces (APIs), they now encompass a wide range of services that are able to run units of work. Ironically, while this work in the scientific community started over a decade ago, it is now the cloud communities that are catching up. As mentioned previously \ref{section:containers}, it has only been in recent years that a need for batch jobs has emerged in Kubernetes, and now these communities are extending this functionality to think of the combination of multiple steps in what we would call a workflow. Batch processing has been an area of expertise for the HPC community for decades \cite{Wikipedia_contributors2023-ha}. This is another opportunity for collaboration and convergence, which can range from the core unit of a batch Job \cite{Section_undated-yk} to an entire operator \cite{Sochat2023-us} deployment. 

Infrastructure engineers are again called to action to advocate for the existing strategies for workflows that have long existed in the HPC community, and find paths for convergence. As an example, Kubernetes APIs might be integrated into existing tools over developing entirely new ones.  An ability to represent DAGs for scientific workflows in the cloud would be an immensely positive development for the scientific community, meaning that we have an ability to extend our workflows to  cloud, and ideally using tools that we already use. The participation of infrastructure engineers in the Kubernetes batch working group meetings will be essential to ensuring this change happens. 

Finally, it must be pointed out that, interestingly, most of the workflow tools to develop in the scientific community are DAG-based. This is logical given the common environments with workload managers to submit work to. However, Kubernetes offers a different model that, although it supports similar batch work, has traditionally been service oriented. An interesting direction for infrastructure engineers to take is to explore the use of state machines for workloads, or rather orchestration of services that respond to changes in state. 

\subsection{Community Participation}
\label{section:possible-future-community}

Most of the paths above, and especially those that require collaboration across cloud and HPC, require addressing the challenge of community participation. From this angle we imagine a possible future where venues for sharing work are attended regularly by both communities. This is starting to happen to some extent. A small number of converged computing practitioners are regularly presenting and speaking at traditionally cloud vendor, or industry conferences such as Kubecon \cite{noauthor_undated-yo,CNCF_Cloud_Native_Computing_Foundation2023-wg,CNCF_Cloud_Native_Computing_Foundation2023-pi}, and advocating for the shared incentives, integration, and co-existence, of technologies and cultures from both spaces. These same community members pushed for and were successful to create the first Kubernetes batch working group that would be friendly to more time zones in the United States.  This trickle of change is also reflected in the venues themselves, in topics related to convergence emerging at the largest HPC conference in the world, Supercomputing \cite{noauthor_undated-an}. Another positive development to send a message to the HPC community is published, collaborative work that spans both spaces \cite{Sochat2024-xd,Misale2022-nc}.

\section{Discussion}

This paper postulates that infrastructure engineers are the core ingredient needed to start much of this work toward a future of continued, successful science. The research community might place greater emphasis on the roles that need to be staffed to afford desired possible futures. While the problems that we face for reproducible and portable scientific workflows might appear superficially at higher levels of the research process, the layer of missing work goes deeper than the scientific software to the core infrastructure and systems themselves. Indeed, we have research software engineers working on components of scientific workflows and domain-specific software to directly address problems in science, but missing are the infrastructure engineers that are oriented to work on paradigm shifts in the larger space of infrastructure -- imagining new futures for scientific workflows in the cloud, and developing the core technologies that are absolutely essential for the continued success of the scientific communities in these new, different environments. A desired future with infrastructure engineers might look like the following:

\begin{itemize}
\item Scientific workflows use paradigms that integrate with cloud technologies
\item Units of portability allow for easy movement between clouds and HPC
\item Previously disparate communities are working together to create better solutions
\item The dichotomy of cloud vs. HPC is non-existent
\item HPC applications can leverage chip and network designs between environments
\item Shared goals are identified, and collaborations common and encouraged
\item Costs are transparent for both cloud and HPC, allowing for informed decisions
\item Resource requests are optimized to fit application needs
\end{itemize}

And importantly, practitioners in the HPC space need to identify priorities for funding and staff, and work backwards to get there. Other facets of these roles are further discussed in the following sections.

\subsection{The importance of creative thinking}

Embedded in the above discussion of challenges and possible futures is an inherent need for creativity and ability to break away from rigid thinking. While stability can be a positive factor for any community to ensure reliability of practices, it can also be a hindrance to a successful adaptation to change, which often is essential for survival. A simple set of questions can be used as a starting point for thinking about any particular technology:

\begin{itemize}
\item How rigid are we in our desire for this to persist?
\item Can we imagine a successful future for it?
\item Can we imagine a successful future without it?
\item How resistant are we to change?
\end{itemize}

These questions hint at cultural and personal patterns of thinking, and require a component of creativity to imagine futures that don't exist yet. Asking these questions can also give insight to the rigidity of one's thinking. A quick response of ``We've done it this way forever'' is a harbinger of this rigidity, or perhaps a fear to diverge from a well-trodden path. Doing so requires creativity and courage on the part of the infrastructure engineer. A prime example of this goes back to the message passing interface (Section \ref{section:mpi}). While it's understandable to want to harness decades of work, a black and white mindset that cannot consider other possible futures where it might work differently or less optimally can actually hinder making progress in the space, and ultimately hurt the scientific community. To extend the example of creative thinking, the following possibilities might be considered for MPI: \newline

\begin{itemize}
\item Change it to work in other environments (e.g., elasticity is needed when resources can be ephemeral and require flexibility).
\item Consider other approaches that will work, initially at some loss of performance, but eventually, maybe not.
\item Demonstrate to cloud that MPI could improve what they are doing, and encourage them to better provide low latency networks.
\item Develop a new cloud-like means to manage processes.
\end{itemize}

All of the approaches above could be reasonable given the right context, and infrastructure engineers to devote time to work on them. In addition to creativity, this work requires persistence and patience. Arguably, the right phenotype of infrastructure engineer might be excited about the innovation and happily jump into the task. Interestingly, although infrastructure engineers must be seasoned software developers, they must also have a strong ability to conduct research. For example, a new idea starts with a proof of concept, then a prototype, and eventually transitions into an experimental setup to provide its worth, and finally, is hardened and used in production. Thinking through possible futures and choosing to work on a subset of them is both a research and a creative task. It also often involves hard work, learning new things, entering states of discomfort, and accepting that the status quo should not always remain that.

\subsection{The importance of collaborative thinking}

The role of infrastructure engineer goes beyond the simple expectation of sitting alongside a research group and writing code. Such a role requires not just innovative and elegant design of software components and infrastructure, but a strong mind for advocacy and presentation of ideas in venues that are not common in academic communities, including industry conferences, online content like blogs, videos, and technical articles, and regular participation in open source communities and standards groups. These engineers must be skilled at identifying areas of developer toil that are often overlooked, and seeing the absence of a cultural or technological component in a large space. Indeed, the skills required for this role are not just technical, but also creative and speculative, with the proper work environment and personality oriented toward taking small risks and enduring many ideas that do not have a fruitful outcome.  

While the academic system has primarily led relationships that are oriented toward teachers and students in a hierarchy of knowledge and power (e.g., professor and graduate student, Principal Investigator and post doctoral candidate), the infrastructure engineer needs to take on a collaborative and collegial role, often needing to work and communicate on the same level as other community members, and drawing most attention to the problems at hand over desires or expectations for prestige. This difference in priorities and culture can often lead to lower stress, or more fun working environments, which is a facet of the work and role that should be emphasized. 

\subsection{Survival of the refactored}

When we consider software and infrastructure as a living entity, it can be stated that software that survives is valued software. Valued software will likely be refactored many times to adapt to a changing environment to run it on. On the flip side, if resources are not adequate to continually work on and improve existing software and underlying interfaces, as the landscape changes a larger set will not survive. While the task of maintaining scientific software likely falls on the research software engineer, the same task for scientific infrastructure falls on the infrastructure engineer. If any set of software that does not survive includes applications that are essential for science, then the loss is very large. Projects that are willing and able to flexibly change with the often changing environments are the ones that will survive. This might even introduce an artificial signal where projects that persist over time (often labeled as being more sustainable) are so because of the mindset of the developers more so than the actual value of the software itself. Likely those two variables are confounded -- valued software will have more developers that consequently keep it updated, but that isn't always the case. 

Finally, it also cannot be overlooked that with paradigm shifts in technology spaces, often there is a shift in jobs and roles needed. With such a shift it's not likely but almost certain that some roles will no longer be needed, and people will need to refocus their expertise for a new scope of work. This is an expected part of a changing landscape that often cannot be avoided, and although the force of change can be stressful or undesired, it often leads to fulfilling roles and overall positive change. It will be up to the community to support smooth shifts between roles, and support of individuals that might struggle with the change.

\subsection{Summary of Innovation Needed}

The following summary of work is provided to the reader as an extended set of explicit suggestions for challenges that also warrant possible futures. \newline

\begin{itemize}
\item \textbf{Transparency for cost and utilization}: This would allow us to improve, better utilize our systems, and make more direct comparisons with cloud.
\item \textbf{Modular developer environments}: Changing our on-premises environments to better afford development of containers and workflows, and capturing patterns of resource usage and application needs would trickle down into innovation for storage and event drive (state machine) architectures.
\item \textbf{Teams of infrastructure engineers}: explicitly to work on HPC infrastructure and technologies. Arguably, many of our current, well established problems result from this missing layer. This area of work includes developing schedulers, standards, workflow tools, and means to assess performance.
\item \textbf{Developer environments}: to afford more flexibility for software versions and provisioning, and capturing of metadata and eventual provenance. This leads to incentives for having better structure and organization of assets.
\item \textbf{System administrator to server ratio}: is improved. Fewer system administrators to manage a larger number of servers would be afforded by more automation. Systems should be easier to manage, and update with fewer non-reoccurring engineering costs (NREs).
\item \textbf{Hardware installation}: is done by companies that exist just to provision it (e.g,. Oxide Computer) and sysadmins are re-allocated to work on software.
\end{itemize}

Not taking measures to innovate, and in a collaborative fashion, poses a huge risk to the scientific community, and (in a worst case scenario) the needs of the community are not met and entire domains of science slowly fade away. For these reasons, it is not just suggested, but imperative, that some of these problems are addressed.

\subsection{Explicit Suggestions for the Reader}

Explicit suggestions are provided to the reader for what an individual can do to have impact. \newline 

\begin{itemize}
\item Develop ``handles'' (language bindings, containers) for adoption across communities. An interested party from a different community is empowered to use the software.
\item Show up to have a voice at the table. This means attending working groups and conferences that are not traditionally attended by your community.  
\item Speak up when a community is not properly represented. One voice can make a difference.
\item Don't criticize or put down something from another community without actively trying to help improve it.
\item Have a collaborative mindset to identify common problems and work toward shared solutions.
\item Don't work within the constraints of role expectations. Figure out what is needed, and find a way to convince others that it should be worked on.
\item Develop applications and infrastructure that is modular, and find analogous counterparts in a different community. 
\item Find incentive structures and policy that affords improving current software and work instead of making something new. 
\item Create software for research that aims to be production quality from the beginning.
\item Find an employer or job role where there is psychological safety to take risks and learn.
\item Ensure there is regular discussion with cloud or similar vendors about plans for software, and hardware. 
\item Participate in standards groups to discuss technology that crosses communities.
\end{itemize}

If it isn't obvious, the above responsibilities are likely too much for a researcher or research software engineer to take on in addition to their current work. This is why the scientific community needs infrastructure engineers -- generalists that are not tied to specific labs or pipelines, but are developing core infrastructure and tools, and acting as community advocates on behalf of the scientific community.

\subsection{Sub-optimal Futures}

The HPC community might run into issues with respect to finding hardware. hardware companies might go out of business if cloud vendors primarily produce hardware in-house. This has bad implications for science, as the HPC community is left with fewer choices, and traditional ways of practicing science start to fail. Niche domains could be taken over by private companies that have handles into the same or different funding sources such as venture capital. Services that are heavily invested in might quickly be deprecated.

Cloud companies, if they are already not, could have hidden monopolies. For example, if one cloud emerges as a leader for providing a specific kind of resource, this could start to look like monopolistic behavior. Special care would be needed to watch out for this, because it can often go unnoticed. An example comes from ice cream brands \cite{Aronczyk2023-vz}. On the ice cream aisle there is an illusion of choice due to the vast array of products to select from. However upon closer inspection, each brand specializes in a specific kind of ice cream (e.g, with and without treats mixed in). This same phenomena could happen with cloud vendors, trapping customers under a single cost model. 

It might also be presented that clouds might get it wrong. More specifically, there is a somewhat unlikely, but not impossible future where the clouds ignore the needs of particular communities, invest all money into one particular niche, and fail to produce revenue. And suddenly, even the option to use cloud is off the table, a still negative shift for the research community. This would still require emergence of new models of compute, either provided commercially or shifted back to the traditional grid computing to meet the needs of science and other practitioners. It could be that then, grid computing providers might need to step back and again consider the strength of their collective resources, and rethink how to harness them, and imagine different models for compute and tenancy. Notably, cloud is taking on a model that does not account for individual users, but rather service roles and namespaces. This would be an interesting thought experiment to apply to high performance computing to possibly allow for a more modular and flexible setup.

The take home point is that if the HPC community is not successful to engage with the cloud community to provide convincing, economic incentive to support the needs of HPC, or if new paradigms are not developed, scientific workloads will not optimally run there, and the science will suffer. If cloud is leading the innovation space, and for both hardware and software, a possible future is that scientists find themselves in a research environment that doesn't support the paradigms that they have committed to.

\section{Conclusions}

This paper points out that in current models for practicing science, not enough attention has been placed on prioritizing the hiring and support of infrastructure engineers to innovate in the space. These developers are accidental or rare, and they are lacking in both resources and institutional backing to be able to successfully do the work that is needed to ensure the continued success of science. 

While an academic mindset for convergence might be scoped to testing ideas and writing papers about it, the task of the infrastructure engineer goes far beyond this expected set of behaviors. A successful and full effort at convergence is about first recognizing and accepting the changing landscape without spite -- namely that cloud is here, leading the space of innovation, dominating the economic landscape, and even hiring talent. In this reality, from the standpoint of an HPC community needing to do science, there are choices to make. The choice can be one of passivity, which ultimately retains the status quo of being left behind. The choice can instead be a task to solve two problems -- to first understand how the needs of science and HPC can be represented in this new environment, and to then to craft a vision for potential futures to enable that. 

These roles, and the challenges that might be solved if they were better supported, are presented transparently to the larger community in this paper. These infrastructure engineers would be the connectors not just between academic institutions, but between industry and their respective institutions. They are the ones that will show up to standards meetings and advocate for scientific communities. They will give talks that aren't about niche research, but the needs and developments of the core technology space. They will be proactive to engage with vendors and industry in a way to ensure the voice of the scientific community is present and heard at the table. This is a call to the scientific community that moving forward, care and attention is taken to not just make room for these roles, but to prioritize them, and likely see the success and innovation that follows that.

\section*{Acknowledgments}
Author VS: to the many individuals that have supported me over the years, often having a vision that did not fit within the confines of an existing community or space. Including my first supervisor, Ruth Marinshaw, that encouraged me to create a position just to work on container technologies and reproducible workflows, to leadership in Livermore Computing that saw my potential to hire me and support me. Thank you to Daniel Milroy for feedback on the manuscript. 

This work was performed under the auspices of the U.S. Department
of Energy by Lawrence Livermore National Laboratory under contract
DE-AC52-07NA27344. Lawrence Livermore National Security, LLC LLNL-JRNL-864868-DRAFT

\bibliographystyle{IEEEtran}
\bibliography{references}

\newpage

\vfill

\end{document}